# Two-dimensional ferromagnetic semiconductors of rare-earth Janus 2H-GdIBr monolayer with large valley polarization

Cunquan Li[1] and Yukai An[1, *]

[1]*Key Laboratory of Display Materials and Photoelectric Devices,
Ministry of Education, Tianjin Key Laboratory for Photoelectric Materials and Devices,
National Demonstration Center for Experimental Function Materials Education,
School of Material Science and Engineering, Tianjin University of Technology, Tianjin, 300384, China*

Two-dimensional ferromagnetic semiconductors with excellent electronic, optical, and magnetic properties have become potential candidates for multiple functional nanodevices. Based on a rare-earth Gd atom with $4f$ electrons, through first-principles calculations, we demonstrate that the Janus 2H-GdIBr monolayer exhibits an intrinsic ferromagnetic (FM) semiconductor character with an indirect band gap of 0.75 eV, high Curie temperature $T_c$ of 260 K, significant magnetic moment of 8 $\mu_B$/f.u. (f.u.=formula unit), in-plane magnetic anisotropy (IMA) and large spontaneous valley polarization of 118 meV. The MAE, inter-atomic distance or angle, and $T_c$ can be efficiently modulated by in-plane strains and charge carrier doping. Under the strain range from −5% to 5% and charge carrier doping from −0.3e to 0.3e/f.u., the system still remains FM ordering and the corresponding $T_c$ can be modulated by strains from 233 K to 281 K and by charge carrier doping from 140 K to 245 K. Interestingly, under various strains, the matrix elements differences ($d_{z^2}$, $d_{yz}$), ($d_{x^2-y^2}$, $d_{xy}$) and ($p_x$, $p_y$) of Gd atoms dominate the MAE behaviors, which originates from the competition between the contributions of Gd-$d$, Gd-$p$ orbitals, and $p$ orbitals of halogen atoms based on the second-order perturbation theory. Inequivalent Dirac valleys are not energetic degenerate due to the time-reversal symmetry breaking in the Janus 2H-GdIBr monolayer. A considerable valley gap between the Berry curvature at the K and K′ points provides an opportunity to selectively control the valley freedom and to manipulate the anomalous Hall effect. External tensile (compressive) strain further increases (decreases) the valley gap up to a maximum (minimum) value of 158 (37) meV, indicating that the valley polarization in the Janus 2H-GdIBr monolayer is robust to the external strains. This study provides a novel paradigm and platform to design the spintronic devices for next-generation quantum information technology.

## I. INTRODUCTION

As described by the Mermin-Wagner theorem [1], the long-range magnetic order of two-dimensional (2D) systems at finite temperature is forbidden by strong fluctuations. Recent studies suggest that magnetic anisotropy opens a gap in the spin-wave spectrum and suppresses the effect of thermal fluctuations. Consequently, when a 2D system exhibits intrinsic anisotropy caused by spin-orbit coupling (SOC), its long-range magnetic order can be well preserved, which has been observed in the CrI₃ monolayer and Cr₂Ge₂Te₆ bilayer with a Curie temperature ($T_c$) of 45 K[2] and 28 K [3], respectively. However, the low $T_c$ is not suitable for the advances of 2D magnetic devices in spintronics. Although, $T_c$ can be moderately increased by carrier doping [4], strain [5], and electric field [6], these extrinsic methods are not ideal for the practical application of devices. Therefore, the exploration of intrinsic 2D ferromagnetic semiconductors with high $T_c$ are urgent for the application of 2D spintronics at nanoscale [7].

Recently, various 2D materials with hexagonal lattice structures, especially for group-VI transition-metal dichalcogenides (TMDs), have opened the possibility of practical implementation of valleytronics. In monolayer TMDs, the conduction-band minima (CBM) and valence-band maxima (VBM) are located at two degenerates but inequivalent valleys at the K and K′ point of hexagonal Brillouin zone due to spatial inversion symmetry breaking [8]. Motivated by this observation, a great deal of interest focuses on exploiting the valley index of so-called K-valley materials, which is rather significant for information processing and storage in electronic devices [9]. If the intrinsic ferromagnetic ordering can introduce spontaneous valley polarization, the namely ferro-valley materials are proposed. Until now, only a few ferro-valley materials have been predicted, including TiVI₆ [10], LaBr₂ [11] and VAgP₂Se₆ monolayers [12], and so on. However, these materials are still not applicable due to the slight intrinsic valley splitting gap and low $T_c$. Most 2D valley materials do not exhibit spontaneous valley polarization due to their temporal inversion symmetry. In general, breaking the inversion symmetry also lifts the spin degeneracy of energy band due to the presence of SOC effect. The introduction of valley polarization is necessary to distinguish and manipulate the carriers in the specified valleys. In other respects, charge carriers can also be distinguished by their spin moments. Thus, many peculiar properties will also be born, such as the quantum valley anomaly Hall effect [13] .However, the results mentioned above more focus on the $d$-electron systems, while the studies on the $f$-electron systems are relatively less. Inspired by the significant magnetic mo-ments and high magneto-crystal anisotropy typically associated with rare-earth elements [14], introducing rare earth elements is expected to largely increase the $T_c$ and magnetization. Very recently, the 2D systems containing Gd with high $T_c$ have been reported, such as GdI$_2$ (300-340 K) [15], GdScSi (352 K) [16] and Gd$_2$B$_2$ (500 K) monolayers [17]. Wang er al. [18] and Feng et al. [19] reported that 2D ferromagnetic GdI$_2$ monolayer exhibits a good dynamic and thermal stability with high $T_c$ of about 241 K. Zhang et al. further proved that the GdCl$_2$ monolayer exhibits the large perpendicular magnetic anisotropy and high $T_c$ (224 K) [20] .However, there are scarce reports about the Gd based-Janus monolayers with the breaking of mirror symmetry. In this work, the stability, electronic, magnetic, and valley properties of rare-earth-based Janus 2H-GdIBr monolayer are investigated systematically. Notably, the Janus 2H-GdIBr monolayer can be easily exfoliated from parent layered bulk crystal and exhibit good thermodynamically and kinetically stability as well as the high $T_c$ of 260 K, large magnetic moment of 8 $\mu_B$/f.u., spontaneous valley polarization of 118 meV and in-plane magnetic anisotropy ($-420$ $\mu eV$/f.u.). The Berry curvature of K and K$'$-valleys shows opposite signs, and the non-zero Berry phase leads to a non-zero anomalous Hall conductivity. These findings strongly suggest the potential application of Janus 2H-GdIBr monolayer in valley electronic devices.

## II. COMPUTATIONAL METHODS

All calculations are performed based on spin-polarized density generalized function theory (DFT) using the Perdew-Burke-Ernzerhof (PBE) function in the generalized gradient approximation (GGA) [21, 22], implemented in the Vienna ab-initio simulation Package (VASP) [23–25]. A vacuum space of 18 Å is set to avoid the interaction between monolayers, and an energy cutoff of 500 eV is set for the plane-wave basis set. The Brillouin zone is sampled using a converged $\Gamma$-centered 8×8×1 k-mesh for structural relaxation and 18×18×1 for the electronic calculations. The standard pseudopotential is used and the valence electron configuration is considered as $5s^2 5p^6 4f^7 5d^1 6s^2$ for Gd [26], $4s^2 4p^5$ for Br, and $5s^2 5p^5$ for I. The crystal structure of Janus 2H-GdIBr monolayer is fully relaxed, and the convergence criteria of energy and forces are $10^{-7}$ eV and 0.005 eV/Å, respectively. The rotationally invariant LSDA+U method is used to handle the strongly correlated corrections to the Gd $4f$ electrons, and the on-site Coulomb interaction parameter (U) and exchange interaction parameter (J) are set at 9.20 and 1.20 eV, respectively [27, 28]. Phonon dispersion spectrum of Janus 2H-GdIBr monolayer is obtained by the PHONOPY code [29, 30] based on the density of functional perturbation theory [31] using a 4×4×1 supercell. The VASPKIT code is used to process the VASP data [32]. Ab initio molecular dynamic (AIMD) simulations are adopted the NVT en-semble based on the Nosé-Hothermostat [33] controlled the temperature of systems at 300 K with a total of 10 ps at 2 fs per step. The Monte Carlo simulation opensource project MCSOLVER [34] is used to estimate the $T_c$ based on the Wolff algorithm of the classical Heisenberg model. 80,000 scans are set to fully thermalize the system to equilibrium within the specified temperature interval starting from the ferromagnetic order. All statistics are obtained from the 720,000 scans that immediately followed. The Berry curvature and anomalous Hall conductivity of Janus 2H-GdIBr monolayer are calculated using maximally localized Wannier functions implemented in the WANNIER90 [35] and WannierTools package [36]. VASPBERRY [37] is used to study its optical properties.

## III. RESULTS AND DISCUSSION

The top and side views of crystal structures for the Janus 2H-GdIBr monolayer are shown in Figure 1a. The Janus 2H-GdIBr monolayer consists of three atomic layers, where the Gd atomic layer is sandwiched between two halogen atomic layers. Meanwhile, the central Gd atom is a trigonal shape coordinated to six halogen atoms. The calculated relaxation equilibrium lattice constants, Gd-I and Gd-Br bond lengths for the Janus 2H-GdIBr monolayer are a=b=4.019 Å, 3.121 Å, and 2.955 Å, respectively. The cleavage energy is calculated from a slab model to confirm the possibility of exfoliation of Janus 2H-GdIBr monolayer from its layered bulk crystal. As shown in Figure 1b, the energy difference increases with the increase of separation distance ($\Delta d = d - d_0$). Eventually, it converges to 0.24 J/m$^2$ at the $\Delta d$=9 Å point, which is taken as the exfoliation energy and confirmed by the calculated cleavage strength (the first derivative of cleavage energy). The exfoliation energy is less than the experimental value of graphite (0.36 J/m$^2$) [38], suggesting the feasibility of exfoliating Janus 2H-GdIBr monolayer experimentally. The calculated magnetic moment of Gd atom is 7.4 $\mu_B$/Gd due to the contribution of half-filled $4f$ and $5d$ shells. Possible magnetic configurations including ferromagnetic (FM) and antiferromagnetic (AFM) states are considered, as shown in Figure 1c, which are the same as those reported previously [20, 39]. To investigate the magnetic ground state and configurations of Janus 2H-GdIBr monolayer, the energies of FM and AFM states of 2×2×1 supercell are compared. The energy of FM state is lower than that of AFM state by 152 meV/Gd [Figure S1], indicating the existence of strong FM coupling. As shown in Figure S2a, with the system always remaining FM state, the variation in the total energy of FM and AFM states can be considered a quadratic function of strains. Charge carrier doping can also effectively modulate the total energy of FM and AFM states [Figure S2b], but the system remains in the FM state. The calculated average electrostatic potential ($\Delta\rho$) along the z-axis is asymmetric with a potential drop of 6.9 eV, as shown in Figure 1d. In the sandwich structure, the potential difference between the two sides implies a change in the work function. Obviously, on the I side, the potential energy is smaller and work function is larger than those on the Br side, which is related to the larger electronegativity of I atom. Moreover, the electrostatic potential difference $\Delta\Phi$ of 0.13 eV can be considered as the redistribution of charges in the Janus 2H-GdIBr monolayer. Figure 1e shows the fluctuation of total energy and snapshots of geometric structures for the Janus 2H-GdIBr monolayer at 10 $ps$ from AIMD simulations and the insets are snapshots of the geometry at 0 $ps$ and 10 $ps$, respectively. Clearly, the simulated total free energy fluctuates within a small range and no significant deformation is observed at the final, suggesting good thermal stability. In addition, the phonon dispersion of Janus 2H-GdIBr monolayer is shown in Figure 1f. The frequencies of all phonon modes within the Brillouin zone are positive, indicating that the Janus 2H-GdIBr monolayer is dynamically stable.

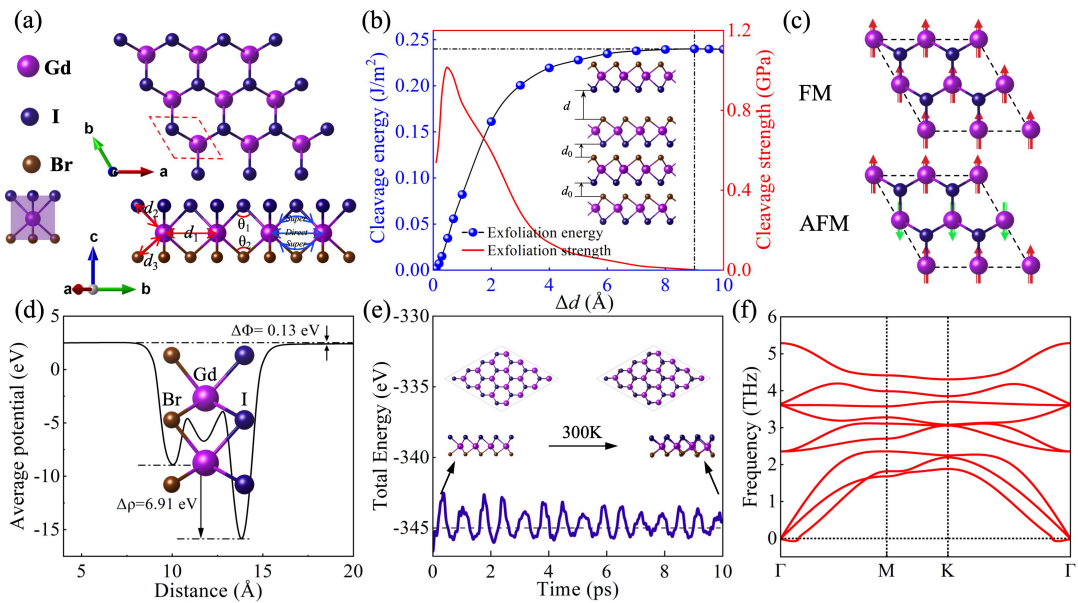

FIG. 1. (a) Top and side views of Janus 2H-GdIBr monolayer with the nearest Gd-Gd distance $d_1$, Gd-I(Br) distance $d_2$ ($d_3$), and Gd-I(Br)-Gd bond angle $\theta_1(\theta_2)$. (b) The dependence of cleavage energy and cleavage strength on the separation distance ($\Delta d = d - d_0$) for the Janus 2H-GdIBr monolayer, where d0 indicates the equilibrium vdW gap in the bulk crystal. The insert shows the calculated exfoliation process. (c) The FM and AFM orders for the Janus 2H-GdIBr monolayer. Red and green arrows denote spin-up and spin-down orientations, respectively. (d) Planar average electrostatic potential energy of Janus 2H-GdIBr monolayer. (e) The fluctuation of total energy and snapshots of geometric structures for the Janus 2H-GdIBr monolayer at 10 $ps$ from AIMD simulations. (f) The phonon dispersion of Janus 2H-GdIBr monolayer.

Figure 2a shows the band structure of Janus 2H-GdIBr monolayer with spin polarization (without considering SOC). The spin splitting leads to an indirect band gap of 0.75 eV, with the VBM at the K(K′) point and the CBM at the M point. Around the Fermi level ($E_F$), the fully spin-polarized valence band and the opposite spin channel conduction band make the Janus 2H-GdIBr monolayer bipolar magnetic semiconductor (BMS). The spin-resolved density of states shown in Figure S3 demon-strates that the $d$ orbitals of Gd atoms contribute to the VBM and CBM for the Janus 2H-GdIBr monolayer. The spin splitting around $E_F$ does not appear for the I and Br atoms, implying that the magnetic moments of Janus 2H-GdIBr monolayer are mainly from the Gd atoms. As shown in Figure 2b, the SOC effect breaks the degeneracy between the K and K′ valleys in the valence band, namely the energy of K valley is higher than that of K′ valley. This results in a large spontaneous valley polarization of 118 meV, which is equivalent to a huge external magnetic field of about 590-1180 T. The valley states at the K and K′ points are mainly contributed by the occupied $d_{x^2-y^2}$ and $d_{xy}$ orbitals of Gd atom and the Γ point is mainly distributed around the $d_{z^2}$ orbital of Gd atom, as shown in Figure S4. The spontaneous valley polarization can be attributed to the strong SOC effect combined with the magnetic exchange interaction of Gd-$d$ electrons due to the breaking of time-reversal symmetry. To further investigate the valley properties, the 3D energy band map in the first Brillouin zone is shown in Figure S5a. By projecting the two energy bands of VBM and CBM onto the horizontal plane, as shown in Figure S5b-c, the energy of VBM at the K and K′ points is not a global but a local minimum, while the energy of CBM at the K and K′ points is a global maximum. The valley polarization of Janus 2H-GdIBr monolayer is much larger than those of other reported ferro-valley materials, e.g., GdF$_2$ (47.6 meV) [40], VAgP$_2$Se$_6$ (15 meV) [12], and LaBr$_2$ (33 meV) [41]. To be a valley material for practical applications, the valley polarization must be large enough. Generally, an estimated valley polarization of 100 meV is required to overcome the thermal noise [42]. In addition, the absence of other bands within the transmission energy window of $-2$ to $0$ eV provides more possibilities for tuning the valley polarization of Janus 2H-GdIBr monolayer, such as applying magnetic fields and changing the magnetic axis orientation.

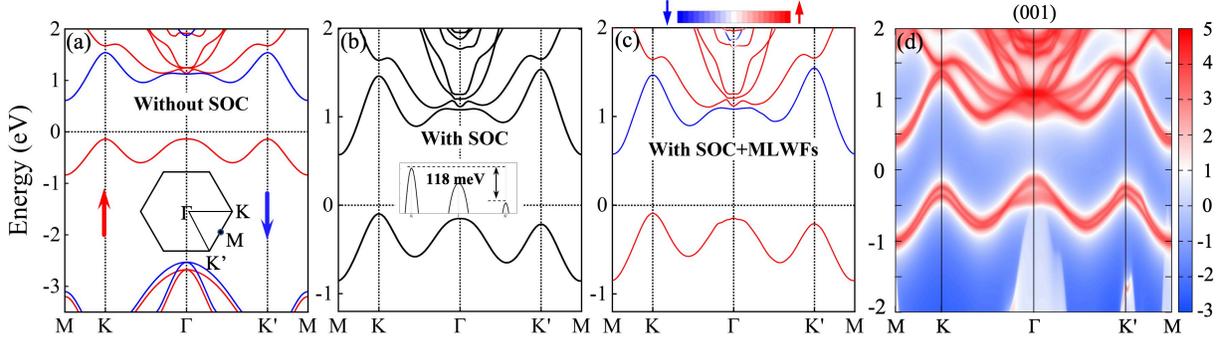

FIG. 2. The band structure of Janus 2H-GdIBr monolayer (a) without SOC; (b) with SOC, and (c) with SOC+MLWFs, respectively. Red (blue) arrows represent the spin up (down). (d) The local density of states of edge states for the Janus 2H-GdIBr monolayer projected on (001) surface. The $E_F$ is set to zero.

To ensure the accuracy of Wannier basis functions, the tightly bound band structure with the spin operator projection ($\hat{S}_z$), obtained from the Wannier interpolation [43] is calculated using MLWFs [Figure 2c], which is basically in full agreement with the DFT results [Figure 2b]. The average spin values of the valence and conduction band edge states are found to be 1 and $-1$, respectively. Thus, their spins remain almost parallel in the upward and downward directions, which usually generates an effective magnetic field that contributes to the energy transfer. Simultaneously, simulated ARPES spectrum [Figure 2d] are calculated using the Wannier function and the iterative Green's function method [44, 45], strongly suggesting that the surface is also a semiconductor state.

Magnetic anisotropy is a critical factor for realizing the long-range FM ordering and the corresponding magnetic anisotropy energy (MAE) can be calculated by considering the transition of magnetic moment from the in-plane [100] to the out-of-plane [001] axis. The Janus 2H-GdIBr monolayer exhibits the in-plane magnetic anisotropy (IMA) with a value of 0.42 meV, which is between the 2H-GdI$_2$ (0.688 meV) and 2H-GdBr$_2$ (0.109 meV) monolayers [20]. Figure 3a shows the MAE of Janus 2H-GdIBr monolayer under various strains. The in-plane strain can be defined as $\varepsilon = (a - a_0)/a_0 \times 100\%$, where $a$ and $a_0$ represent the in-plane lattice constants of strained and unstrained Janus 2H-GdIBr monolayer, respectively. It is obvious that the MAE of Janus 2H-GdIBr monolayer shows a monotonous increase as the strain increases from $-8\%$ to $8\%$. The atomic layer-resolved MAEs for the Janus 2H-GdIBr monolayer under various strains are shown in Figure 3b. One can see that the MAE of Janus 2H-GdIBr monolayer is mainly determined by the Gd and I atoms. The Br and I atoms show the IMA character, while the Gd atom exhibits a transition from the PMA to IMA characters at the strain increases from $-8\%$ to $8\%$. The orbital-resolved MAE can be defined by using the second-order perturbation theory:

$$\text{MAE} = \xi^2 \sum_{o,u} \frac{\left| \left\langle \psi_o \left| \hat{L}_x \right| \psi_u \right\rangle \right|^2 - \left| \left\langle \psi_o \left| \hat{L}_z \right| \psi_u \right\rangle \right|^2}{E_u - E_o} \quad (1)$$

where $E_o$ ($E_u$) represent the energies of occupied (non-occupied) states and the $\xi$, $o$, and $u$ are the SOC constants. The Gd-d, Gd-p and I(Br)-p orbital resolved MAEs of Janus 2H-GdIBr monolayer at the strains of $-6\%$, $0\%$, and $6\%$ are shown in Figure 3c and Figure S6. For the unstrained (0%) Janus 2H-GdIBr monolayer, more significant positive MAE (IMA) can be attributed to the matrix elements differences ($p_x$, $p_y$) of I atom as well as the matrix elements differences ($d_{z^2}$, $d_{yz}$) and ($p_x$, $p_y$) of Gd atom. When the strain changes from $-6\%$ to $6\%$, the positive MAE increases, which is mainly caused by the increase of matrix element difference ($d_{z^2}$, $d_{yz}$) of Gd atom and the decrease of matrix element difference ($p_x$, $p_y$) and ($d_{x^2-y^2}$, $d_{xy}$) of Gd atom. It is noted that the matrix elements difference ($d_{z^2}$, $d_{yz}$) and ($d_{x^2-y^2}$, $d_{xy}$) of Gd atom shows the opposite signs at the strain of $-6\%$. However, at the strain of 6%, the matrix element difference ($d_{x^2-y^2}$, $d_{xy}$) of Gd atom remarkably decreases, which is weak to the matrix element difference ($d_{z^2}$, $d_{yz}$) of Gd atom. Meanwhile, the matrix elements differences ($p_x$, $p_y$) of I and Br atoms keep stable and the matrix elements differences ($p_x$, $p_y$) of Gd atom monotonously decrease, which results in the appearance of IMA character. Thus, the matrix elements differences ($d_{z^2}$, $d_{yz}$), ($d_{x^2-y^2}$, $d_{xy}$) and ($p_x$, $p_y$) of Gd atoms dominate the MAE behaviors under various strains for the Janus 2H-GdIBr monolayer. The dependence of MAE (MAE=$E^{[001]} - E^{\theta}$) on the polar angle ($\theta$) is also investigated. Figure 4a shows the angular dependence of MAE along the X (100), Y (010), and Z (001) axes. It is clear that the MAE strongly depends on the direction of magnetization in the $xz$ and $yz$ planes. In contrast, the MAE in the $xy$ plane is isotropic. Thus, a strong dependence of MAE on the angle of magnetization in the out-of-plane is observed, which is similar to the case of VS$_2$ [47] and FeCl$_2$ [48] monolayers. The strong magnetic anisotropy of Janus 2H-GdIBr monolayer can be confirmed by the distribution of MAE on the whole space, as shown in Figure 4b. The MAE is zero in the xy-plane and reaches a maximum value of 415 $\mu$eV Gd$^{-1}$ in the $xz$ ($yz$) plane, which is comparable to the GdI$_2$ monolayer ( 553 $\mu$eV Gd$^{-1}$) [20].

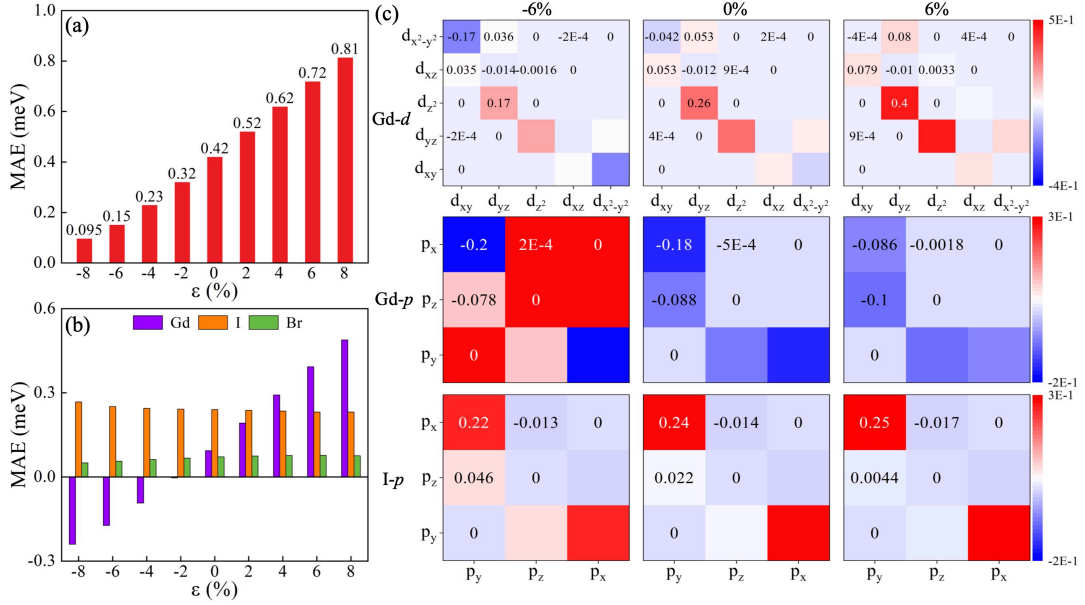

FIG. 3. (a) Total and (b) atomic-layer-resolved MAEs for the Janus 2H-GdIBr monolayer under various strains. (c) Gd-d, Gd-p and I-p orbital resolved MAEs for the Janus 2H-GdIBr monolayer under biaxial strains of −6%, 0% and 6%.

High T$_c$ is crucial for the practical application of spintronic devices. To determine the long-range magnetic exchange interactions of Janus 2H-GdIBr monolayer, the metropolis Monte Carlo algorithm based on the Heisenberg model is used, and the spin Hamiltonian can be described as [49]:

$$H = -J \sum_{i,j} S_i S_j - D \sum_i \left(S_i^z\right)^2 \qquad (2)$$

where $S_i$ and $S_j$ are the spin operator, $J$ represents the isotropic part of exchange interaction, and $D$ represents the single-ion magnetic anisotropy on iron. All parameters in Eqn. (1) are determined by the energy mapping in relativistic density generalized function theory calculations for the $2 \times 2 \times 1$ supercells, including cells with the FM and AFM configurations. The energies of FM and AFM states and the magnetic ion anisotropy parameter $D$ with considering the SOC effect can be calculated as follows:

$$E_{FM} = E_0 - 6J|S|^2 - D|S|^2 \qquad (3)$$

$$E_{AFM} = E_0 + 2J|S|^2 - D|S|^2 \qquad (4)$$

where $E_{FM}$ and $E_{AFM}$ are the total energies of FM and AFM configurations. $E_0$ is the total energy in the system without magnetic exchange coupling. $|S|$ is the total spin of Gd. The nearest neighbor exchange parameter $J$ is defined as:

$$J = \frac{E_{AFM} - E_{FM}}{8|S|^2} \qquad (5)$$

The calculated $J$ value for the Janus 2H-GdIBr monolayer is 1.188 meV, suggesting that the magnetic interactions between the nearest neighboring Gd atoms are in FM state. Figure 4c shows the temperature dependences of average magnetic moment and speciated heat for the Janus 2H-GdIBr monolayer. During the simulation from 0 K to 400 K, T$_c$ can be estimated to be about 260 K. Also, using the same method, the calculated T$_c$ of CrI$_3$ monolayer is about 47 K, which is similar to the experimental result of 45 K [50], proving the reliability of the results. Combined with the calculated high T$_c$, the Janus 2H-GdIBr monolayer can be expected as an ideal ferrovalley material for applications in valley electronics. A real-space renormalization group analysis is used to avoid errors originating from finite size by roughly dividing the original 32×32 lattices in a 16×16 lattice, with every four adjacent spins forming quasiparticles, each with a representative spin. The Heisenberg Hamiltonian of the renormalized model is assumed to be the same as before, and the internal energy of this system is recalculated. More details are shown in Figure S7.

The highly localized $4f$ electrons on Gd atoms is negligible in the magnetic exchange coupling. However, more extended $5d$ electrons on Gd atoms occupy the spin-up channel near the $E_F$, making the Janus 2H-GdIBr monolayer a magnetic semiconductor. According to the Kramer mechanism, the partially occupied states can result in the direct exchange interaction between the nearest Gd-Gd atoms, and the bond angle of Gd-I(Br)-Gd is close to 90°, satisfying the Goodenough-Kanamori-Anderson (GKA) rules [51–53]. To investigate the origin of FM coupling in the Janus 2H-GdIBr monolayer, interatomic interactions need to be considered. Generally, the direct (super) exchange depends on the Gd-Gd (Gd-I(Br)-Gd) bond, which leads to the FM (AFM) states. Generally, the Gd-Gd (I/Br) distance and the Gd-I(Br)-Gd bond angle mainly contribute to this magnetic exchange mechanism. Thus, the synergistic effects of direct and super exchange interactions determine the magnetic ground state of Janus 2H-GdIBr monolayer. The high concentration of hole carrier doping usually induces a transition from the AFM to FM states, proving the effectiveness of modulating the magnetic behavior by charge carrier doping. As shown in Figure 5a-b, the nearest Gd-Gd distance $d_1$ always remains constant, and the Gd-I distance $d_2$, and Gd-Br distance $d_3$ show a monotonical and slight increasing (decreasing) trend under electron (hole) doping. Meanwhile, the Gd-I-Gd bond angle $\theta_1$/Gd-Br-Gd bond angle $\theta_2$ increases (decreases) with increasing hole (electron) concentration, which significantly strengthens (weakens) the super-exchange interaction. During the experiment, various strains caused by the lattice mismatch of substrate is common. Figure 5c-d shows the dependence of the Gd-Gd(I/Br) distance $d$ and the Gd-I(Br)-Gd bond angle $\theta$ on strains. As the strain increases from $-8\%$ to 8%, the nearest Gd-Gd distance $d_1$ (from 3.70 Å to 4.34 Å) and Gd-I-Gd bond angle $\theta_1$ (from 74.2° to 85.5°)/Gd-Br-Gd bond angle $\theta_2$ (from 79.4° to 91.0°) exhibit the monotonical increasing behaviors. According to the GKA rule, the increased (decreased) $d_1$ weakens (strengthens) the direct exchange interaction. Meanwhile, the increased (decreased) angle $\theta_1/\theta_2$ enhances (weakens) the super-exchange inter-action. Clearly, the direct exchange and super-exchange interactions compete with each other. Combing with the dependence of exchange parameter $J$ on strain [Figure S8a], the increased or decreased $d_1$ makes the direct exchange interaction more critical in determining the FM state. Generally, the tensile (compressive) strain weakens (enhances) the FM coupling and plays a vital role in the direct and super exchange mechanisms. Meanwhile, in the range of hole doping, the exchange parameter $J$ changes slightly, and the $T_c$ near room-temperature is well preserved for the Janus 2H-GdIBr monolayer. In contrast, the exchange parameter $J$ and corresponding $T_c$ change significantly with increasing electron doping [Figure S8b-d]. The tensile (compressive) strain tends to monotonically decrease (increase) the exchange parameter $J$ and corresponding $T_c$. Interestingly, the maximum $T_c$ reaches to 281 K at the tensile strain of 8%, close to the room temperature [Figure S8a, c-d].

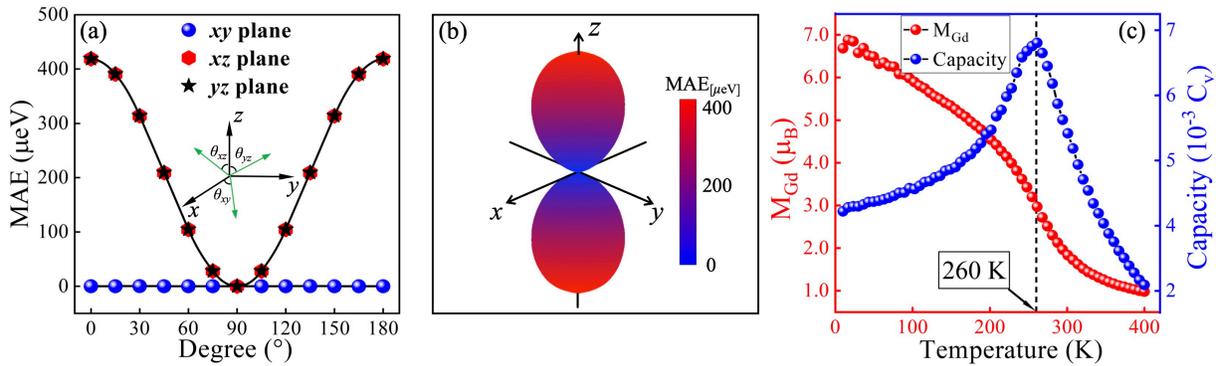

FIG. 4. Dependence of MAE on the polar angle ($\theta$) for the Janus 2H-GdIBr monolayer with the direction of magnetization lying on (a) $xy$, $xz$ and $yz$ planes as well as (b) the whole space. (c) Evolution of average magnetic moment and specific heat for the Janus 2H-GdIBr monolayer.

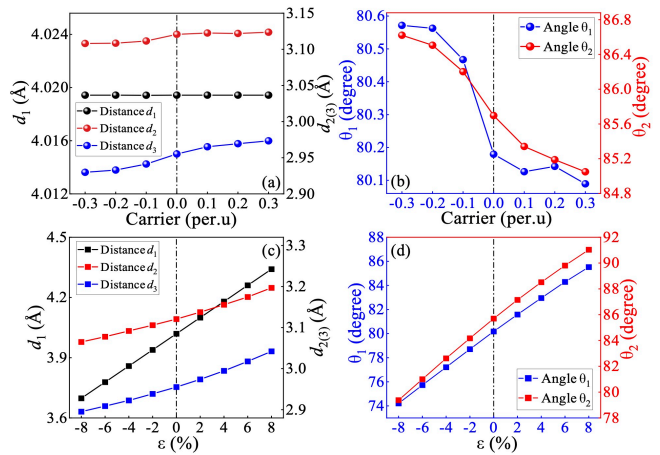

FIG. 5. Charge carrier doping dependence of the inter-atomic (a) distance and (b) angle for the Janus 2H-GdIBr monolayer. Biaxial strain dependence of the inter-atomic (c) distance and (d) angle for the Janus 2H-GdIBr monolayer.

Berry curvature is closely related to the Hall effect of

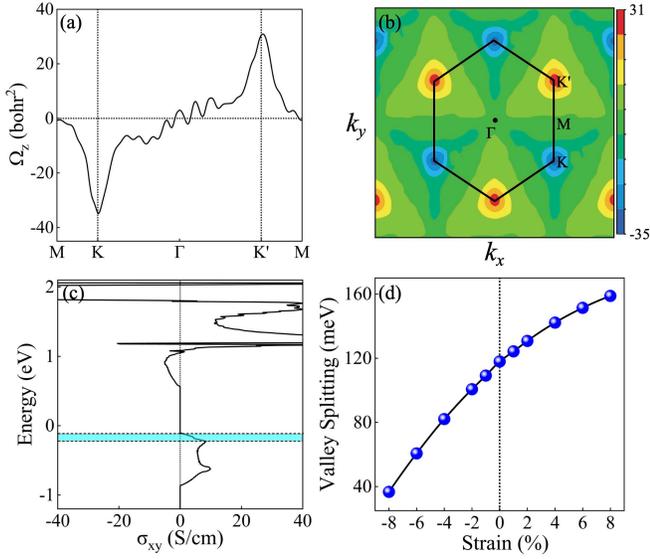

FIG. 6. Calculated Berry curvature of Janus 2H-GdIBr monolayer (a) along high symmetry lines and (b) over the 2D Brillouin zone. (c) Calculated anomalous Hall conductivity of Janus 2H-GdIBr monolayer, the two dashed lines denote the two valley extrema. (d) Variation of valley polarization for the Janus 2H-GdIBr monolayer as a function of strain from $-8\%$ to $8\%$.

the system. When the space inversion symmetry is broken in the hexagonal systems, the charge carriers at the K and K′ valleys can generate the non-zero Berry phases along the z-direction in accompany with a non-zero Berry curvature. When the time-reversal symmetry is also broken, the characteristic of valley contrast appears. To investigate this property, the Berry curvature of Janus 2H-GdIBr monolayer is calculated according to the Kubo formula [54]:

$$\Omega(k) = -\sum_n \sum_{n \neq m} f(n) \frac{2 \operatorname{Im} \langle \varphi_{nk} | v_x | \varphi_{mk} \rangle \langle \varphi_{mk} | v_y | \varphi_{nk} \rangle}{(E_n - E_m)^2}$$ (6)

where $v_x$ and $v_y$ are velocity operators of the Dirac electrons along $x$ and $y$ directions, respectively. $f(\mathrm{n})$ is the Fermi-Dirac distribution function for the $n^{th}$ band, and $|\varphi_{nk}\rangle$ is the calculated Bloch wave function with the energy eigenvalue $E_n(E_m)$. The maximally localized Wannier function is used to calculate the Berry curvature. Figure 6a-b shows the Berry curvature of Janus 2H-GdIBr monolayer along the high symmetry line and over the 2D Brillouin zone. It is clear that the K and K′ valleys have opposite signs of Berry curvature and slightly different absolute values, revealing that the valley contrast phenomenon plays an important role in characterizing the Bloch electron chirality in k-space [55]. Away from the K and K′ valleys, Berry curvature decays rapidly and disappears at the M point. As described in Eqn. (6) [56], the Berry curvature drives a peculiar transverse velocity

in the presence of an in-plane electric field $E$:

$$v = -\frac{e}{\hbar} E \times \Omega(k)$$ (7)

This is an intrinsic contribution of the anomalous Hall effect. The integral of Berry curvature in the Brillouin zone gives the contribution to the anomalous Hall conductivity (AHC), which can be defined as [57]:

$$\sigma_{xy} = -\frac{e^2}{\hbar} \int_{BZ} \frac{d^2 k}{(2\pi)^2} \Omega(k)$$ (8)

Due to the opposite sign and unequal absolute value of Berry curvature, the charge carriers in the K and K′ valleys have opposite transverse velocities. Thus, anomalous Hall conductivity with a net value of non-zero is generated. As shown in Figure 6c, when the $E_F$ is located between the valence band edges of the K and K′ valleys, a fully valley-polarized Hall conductivity is generated with a calculated maximum AHC of 9.5 S/cm. In addition, the $E_F$ can be effectively tuned above (below) the valley splitting gap by a low concentration of carrier doping [Figure S9]. On the other hand, if a finite valley polarization can be generated, for example, by irradiating with circularly polarized light [see details in Figure S10], the Hall currents will also appear. As shown in Figure 6d, the valley polarization increases monotonically with the increase (decrease) of tensile (compressive) strains. And a giant valley polarization of about 158 meV is obtained at tensile strain of 8%. But the valley polarization remains significant (about 37 meV) under a considerable compressive strain of $-8\%$, implying that the valley polarization is robust against the in-plane strain. Therefore, the Janus 2H-GdIBr monolayer can be considered as an ideal ferro-valley material providing a potential platform for valley electronic applications.

## IV. CONCLUSIONS

In summary, by first principles calculations, we predict that the Janus 2H-GdIBr monolayer exhibits a low exfoliation energy of 0.24 J/m² and possesses tunable semiconductor character with the sizeable magnetic moment, high $T_c$ of 260 K, in-plane magnetic anisotropy, and excellent thermal and dynamic stability. The MAE and valley polarization features are robust against the in-plane biaxial strains. The total MAE can increase monotonically from nearly 0.1 to 0.81 meV/f.u. under strain from $-8\%$ to 8%. Meanwhile, the easy axis of Gd atom show a transition from the [001] to [100] direction. Based on the second-order perturbation theory, it is found that the competition between the contributions of Gd-$d$, Gd-$p$ orbitals, and $p$ orbitals of halogen atoms induce the IMA character. When the tensile strain of 8% is applied, the valley polarization of Janus 2H-GdIBr monolayer can reach a maximum value of 158 meV. Furthermore, the corresponding $T_c$ is tuned by strains from 233

K (8%) to 281 K ($-8\%$) and charge carrier doping from 140 K ($0.3e$/f.u.) to 245 K ($-0.3e$/f.u.). In addition, the origin of FM coupling in the Janus 2H-GdIBr monolayer can be attributed to the direct-exchange (Gd-Gd) and super-exchange (Gd-I(Br)-Gd) interactions. Due to the space-inversion and time-reversal symmetry breaking, an appropriate external electric field can achieve the anomalous valley Hall effect in the Janus 2H-GdIBr monolayer.

Overall, it can be expected that Janus-2H-GdIBr monolayer is a new candidate for the next generation of spintronic and valleytronic devices.

## ACKNOWLEDGMENTS

This work was supported by Natural Science Foundation of Tianjin City (Grant No. 20JCYBJC16540).

# Two-dimensional ferromagnetic semiconductors of rare-earth Janus 2H-GdIBr monolayer with large valley polarization

Cunquan Li[1], Yukai An[1,*]

[1] Key Laboratory of Display Materials and Photoelectric Devices,
Ministry of Education, Tianjin Key Laboratory for Photoelectric Materials and Devices,
National Demonstration Center for Experimental Function Materials Education,
School of Material Science and Engineering, Tianjin University of Technology, Tianjin, 300384, China

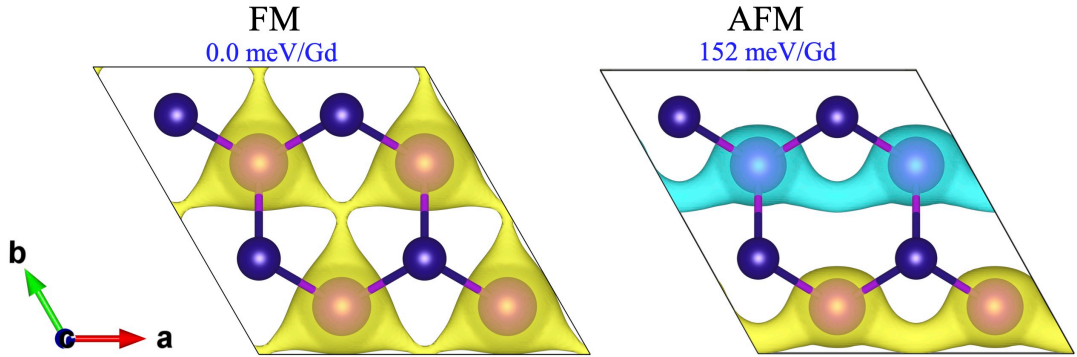

FIG. S1. Spin density of 2×2×1 Janus 2H-GdIBr monolayer for different magnetic configurations. Yellow (blue) color refers to the spin up (down) charge density. The isosurface is 0.0094 e/Born$^3$.

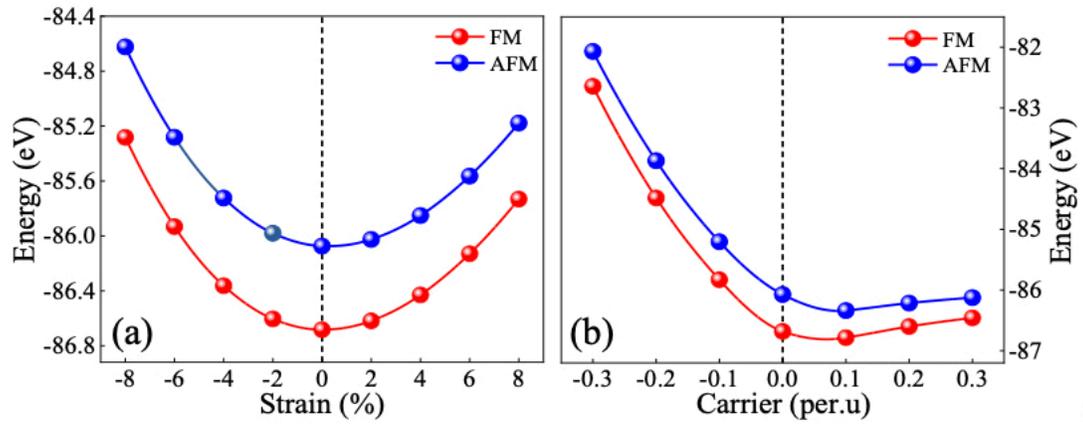

FIG. S2. (a) Strain and (b) charge carrier dependence of the energy in FM and AFM states for the 2×2×1 Janus 2H-GdIBr monolayer.

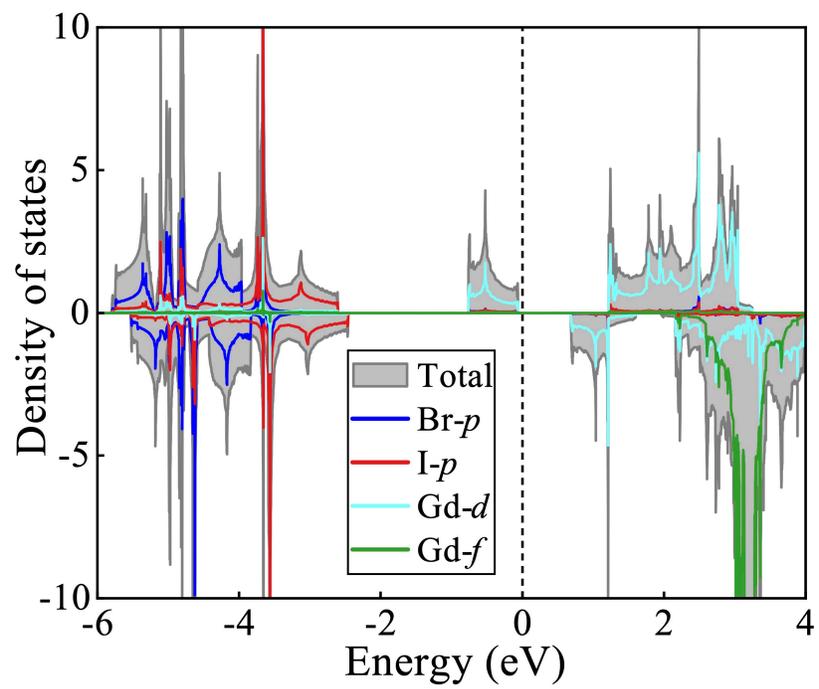

FIG. S3. The spin-resolved PDOSs of Janus 2H-GdIBr monolayer.

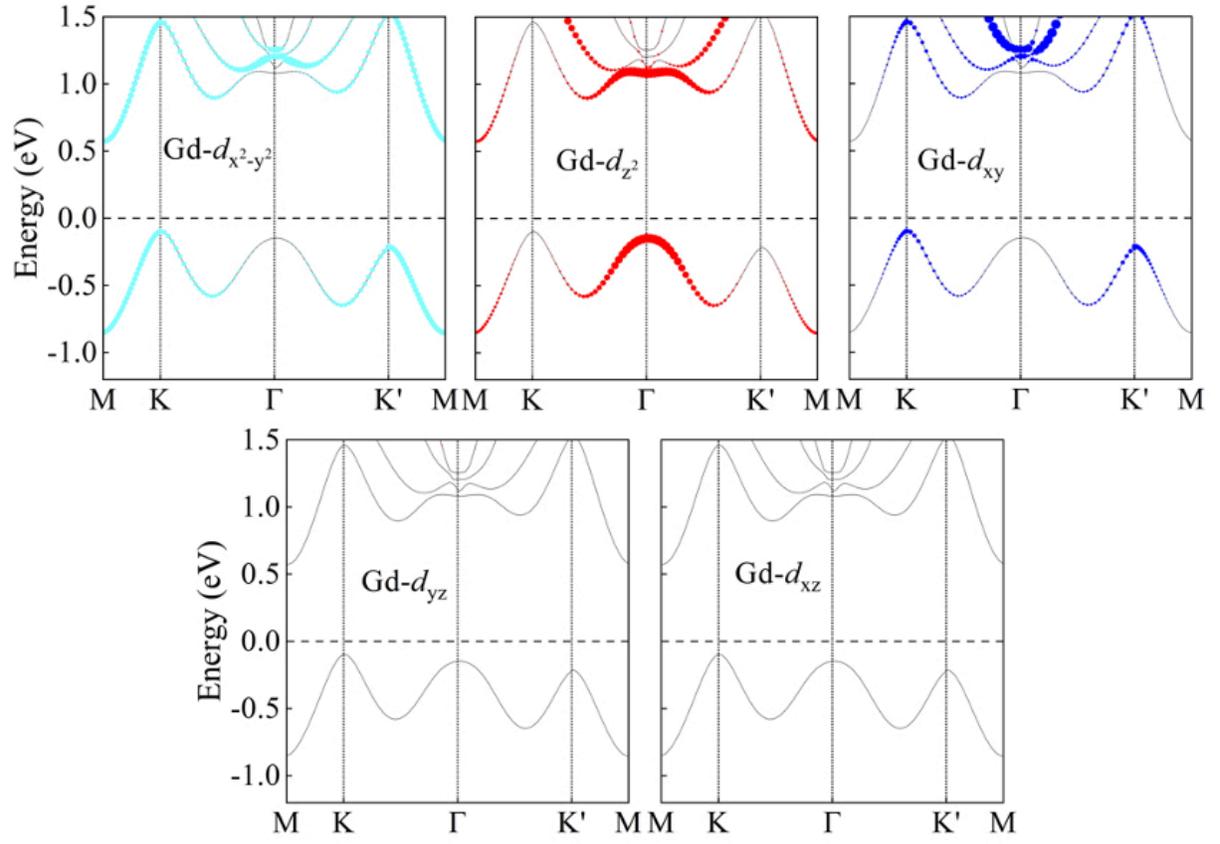

FIG. S4. Projected band structure of Gd-$d$ orbitals for the Janus 2H-GdIBr monolayer with SOC. The Fermi level is set to zero.

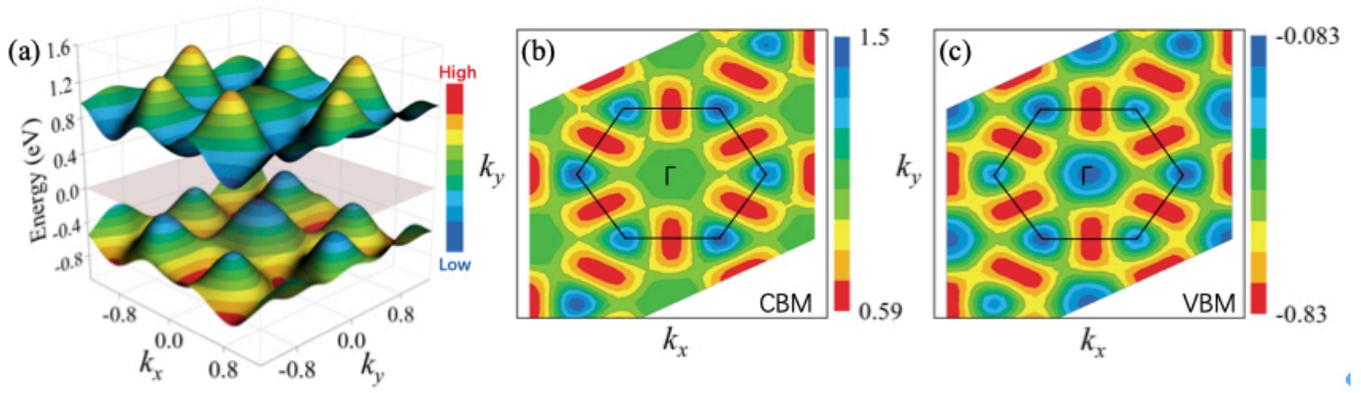

FIG. S5. (a) 3D band structure as well as the 2D projected band structure of (b) CBM and (c) VBM at the $k_x k_y$-plane for the Janus 2H-GdIBr monolayer with considering SOC. The different colors in color bar show different iso-values. The blue and red color correspond to the smallest and largest energy values, respectively.

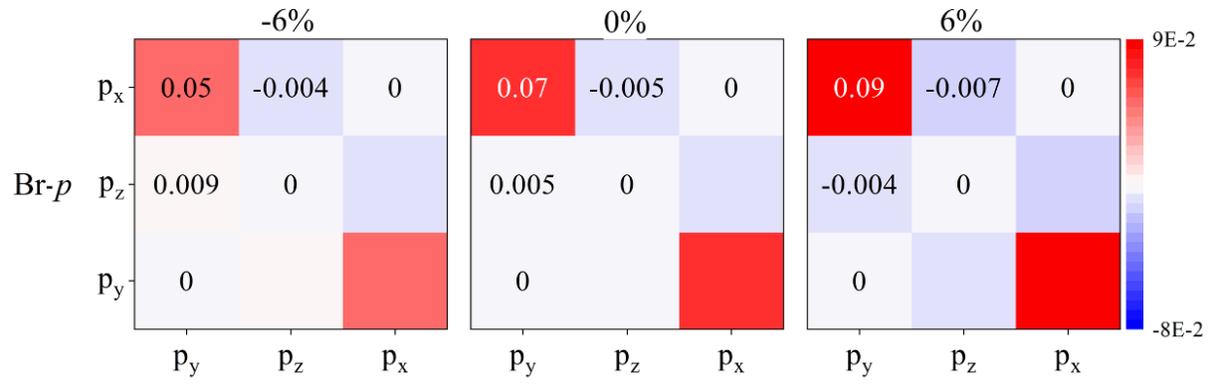

FIG. S6. Br-p orbital resolved MAEs for the Janus 2H-GdIBr monolayer at the strains of −6%, 0% and 6%.

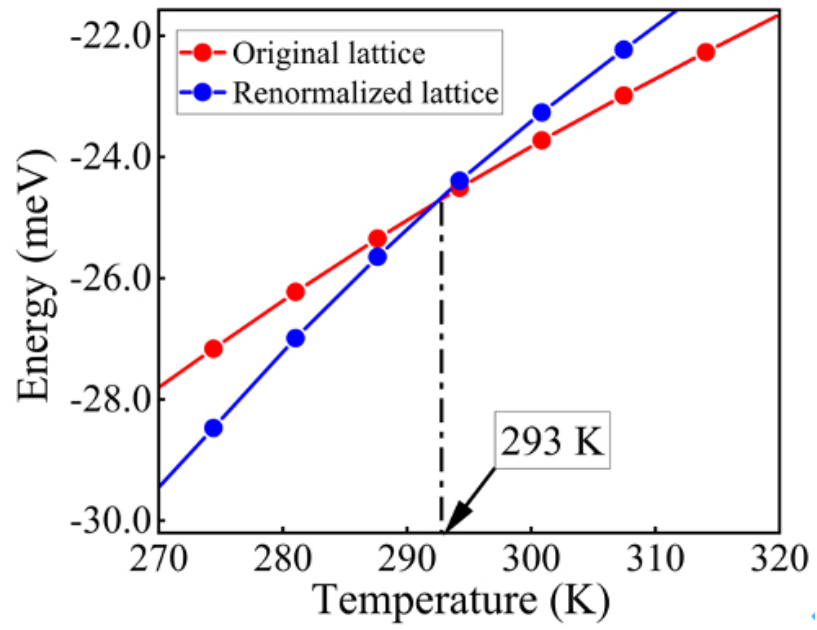

FIG. S7. Nearly linear evolution of the internal energy per unit cell in the original 16×16 superlattice of spins (red) and the renormalized 32×32 superlattice of quasi-particles (blue). Lines are only to guide the eye.

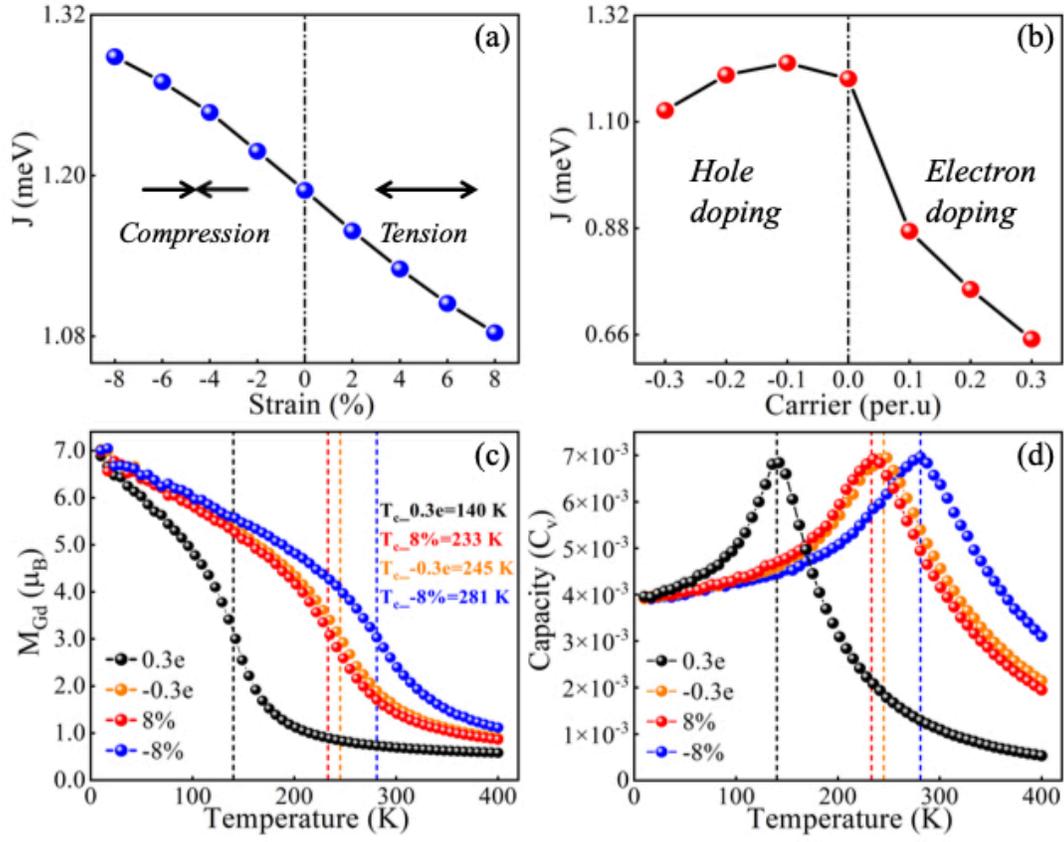

FIG. S8. Dependences of exchange parameter $J$ on (a) biaxial strain and (b) charge carrier doping for the Janus 2H-GdIBr monolayer. Dependences of (c) magnetic moment and (d) heat capacity on the temperature for the Janus 2H-GdIBr monolayer at the strains of $\pm 8\%$ and with carrier concentration of $\pm 0.3$e per.u.

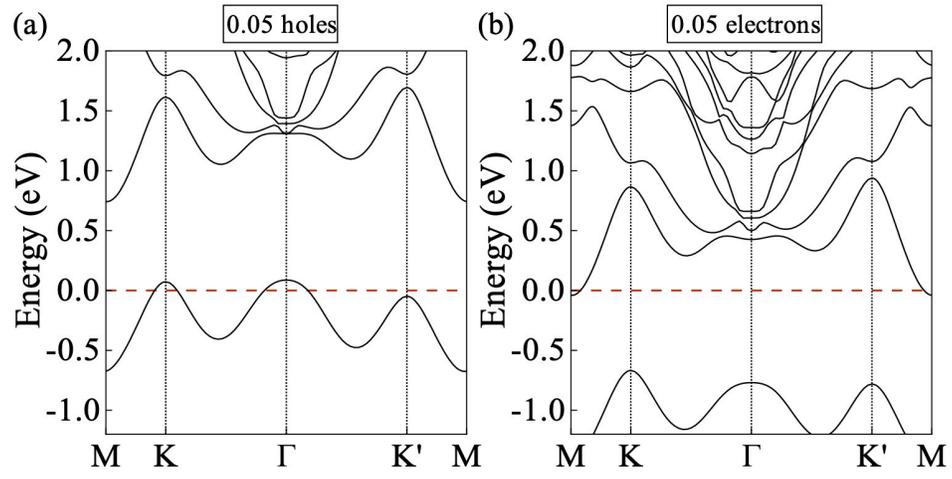

FIG. S9. The band structures Janus 2H-GdIBr monolayer with (a) 0.05 holes and (b) 0.05 electrons doping. The Fermi level is set to zero.

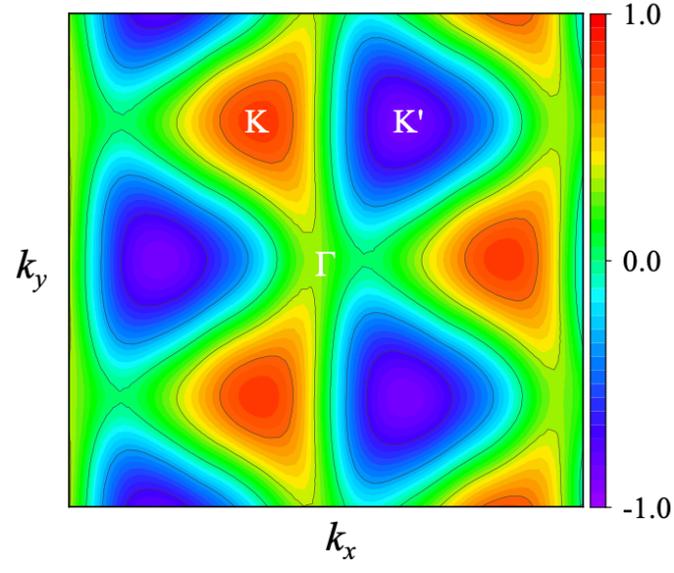

FIG. S10. The circular polarization $\eta(\mathrm{k})$ of the optical transition between the VBM and CBM in the first Brillouin zone.